\documentclass[aps,prb,twocolumn,superscriptaddress]{revtex4-1}
\usepackage{amssymb, amsmath}
\usepackage{graphicx,onlyamsmath}

\usepackage{natbib}

\begin{document}

\title{Pressure and temperature driven phase transitions in HgTe quantum wells}

\author{S.~S.~Krishtopenko}
\affiliation{Institute for Physics of Microstructures RAS, GSP-105, 603950, Nizhni Novgorod, Russia}
\affiliation{Laboratoire Charles Coulomb, UMR CNRS 5221, University of Montpellier, 34095 Montpellier, France.}

\author{I.~Yahniuk}
\affiliation{Institute of High Pressure Physics, Polish Academy of Sciences, Soko{\l}owska 29/37 01-142 Warsaw, Poland.}

\author{D.~B.~But}
\affiliation{Laboratoire Charles Coulomb, UMR CNRS 5221, University of Montpellier, 34095 Montpellier, France.}

\author{V.~I.~Gavrilenko}
\affiliation{Institute for Physics of Microstructures RAS, GSP-105, 603950, Nizhni Novgorod, Russia.}
\affiliation{Lobachevsky State University of Nizhni Novgorod, pr. Gagarina 23, 603950, Nizhni Novgorod, Russia.}

\author{W.~Knap}
\affiliation{Laboratoire Charles Coulomb, UMR CNRS 5221, University of Montpellier, 34095 Montpellier, France.}
\affiliation{Institute of High Pressure Physics, Polish Academy of Sciences, Soko{\l}owska 29/37 01-142 Warsaw, Poland.}

\author{F.~Teppe}
\email[]{frederic.teppe@umontpellier.fr}
\affiliation{Laboratoire Charles Coulomb, UMR CNRS 5221, University of Montpellier, 34095 Montpellier, France.}

\date{\today}

\begin{abstract}
We present theoretical investigations of pressure and temperature driven phase transitions in HgTe quantum wells grown on CdTe buffer. Using the 8-band \textbf{k$\cdot$p} Hamiltonian we calculate evolution of energy band structure at different quantum well width with hydrostatic pressure up to 20 kBar and temperature ranging up 300 K. In particular, we show that in addition to temperature, tuning of hydrostatic pressure allows to drive transitions between semimetal, band insulator and topological insulator phases. Our realistic band structure calculations reveal that the band inversion under hydrostatic pressure and temperature may be accompanied by non-local overlapping between conduction and valence bands. The pressure and temperature phase diagrams are presented.
\end{abstract}

\maketitle

\section{\label{sec:Introduction}Introduction}
Almost a decade ago a new class of materials, so-called topological insulators (TI), was predicted\cite{w1,w2}. TIs possess a band gap for the bulk states and gapless edge states. These edge states are protected against single-particle perturbations by time reversal symmetry.\cite{w1,w2,w3,w4,w5,w6} TI systems can be found in materials, in which the conduction and valence bands have opposite parity and a change in the band ordering occurs\cite{w4}.

\begin{center}
\begin{figure}[h]
\includegraphics [width=1.0\columnwidth, keepaspectratio] {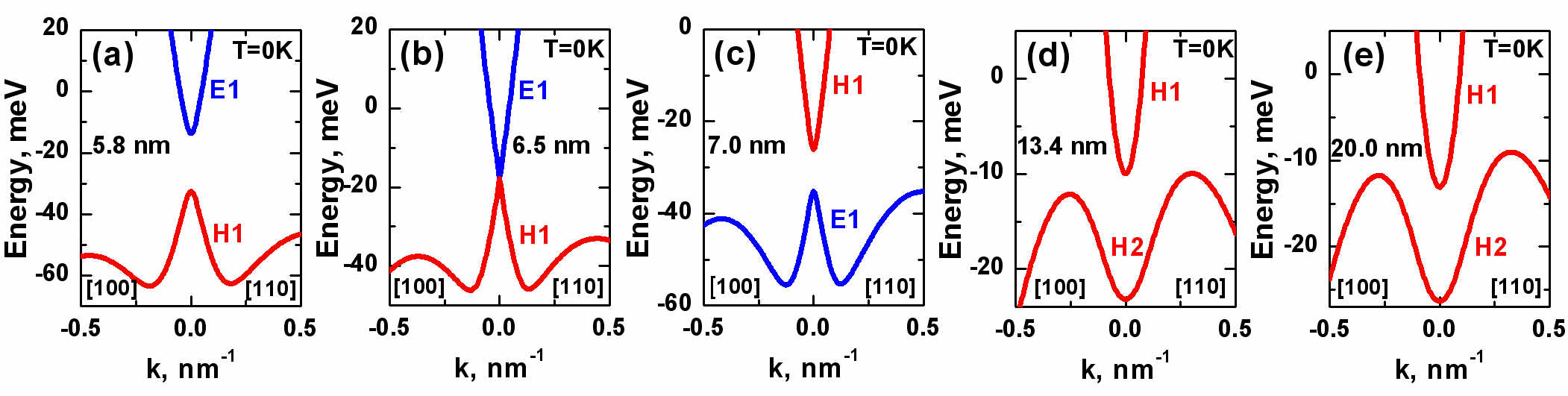} 
\caption{\label{Fig:1} (Color online) Typical band structure of (001)-oriented HgTe QWs at zero temperature and at different QW width: (a) BI phase, $d<d_c$, (b) Dirac cone, $d=d_c$, (c) TI phase $d>d_c$, (d) gapless state, $d=d_{SM}$, (d) SM phase, $d>d_{SM}$. Electron-like E1 subband is shown in blue, while red curves correspond to the heavy-hole subbands. In the panels (d) and (e), E1 subband lies significantly lower in energy. Here, we do not differ TI state, in which the band gap is defined by the gap between H1 and E1 subband, and TI states, in which the E1 subband lies below several heavy-hole-like subbands.\cite{w31}}
\end{figure}
\end{center}

The first TIs discovered were based on HgTe/Cd(Hg)Te quantum wells (QWs).\cite{w4,w5} This two-dimensional (2D) system can be tuned from the trivial band insulator (BI) to the 2D TI phase by changing the QW width $d$. The origin of 2D TI phase is caused by the inverted band structure of HgTe, which leads to a peculiar size quantization in HgTe/Cd(Hg)Te QWs. Specifically, as $d$ is varied, the lowest 2D subband, formed by coupling of conduction band ($\Gamma_6$) states with light-hole ($\Gamma_8$) states and defined as electron-like level (E1 subband), crosses the top subband of heavy-hole ($\Gamma_8$) states (H1 subband)\cite{w4}. When $d$ exceeds the critical width $d_{c}$ the E1 subband falls below the H1 subband and the 2D system has inverted band structure (see Fig.~\ref{Fig:1}c). The critical width also depends on the crystallographic orientation and buffer material, on which the QW is grown. For HgTe/Cd$_{0.7}$Hg$_{0.3}$QWs grown on CdTe buffer, $d_{c}\approx6.5$ and 6.3 nm for (001) and (013) orientations respectively. In narrow HgTe QWs ($d<d_c$), a conventional alignment of electronic states (CdTe-like) with BI phase can be obtained (see Fig.~\ref{Fig:1}a). Thus, a topological phase transition occurs at the critical thickness $d_c$, at which the band gap is absent and the system is characterized by the linear dispersion of massless Dirac fermions.\cite{w12}

In wide HgTe/Cd(Hg)Te QWs the side maxima of the valence band overlaps with the conduction band (see Fig.~\ref{Fig:1}e), while $d=d_{SM}$ corresponds to the indirect gapless state (Fig.~\ref{Fig:1}d). A semimetal (SM) phase at $d>d_{SM}$ is then formed when the Fermi level crosses both the valence and conduction bands.\cite{w13,w13a,w15} Recent finding proves\cite{w13,w13a} that SM phase is a universal property of wide HgTe QWs independent of the surface orientation. Additionally, the overlapping of the valence and conduction bands, thus the SM phase, is very sensitive to the strain effects\cite{w13a} caused by the lattice mismatch of HgTe and CdTe.

A reliable fingerprint of the band inversion is the characteristic behavior of a particular pair of Landau levels (LLs), so-called \emph{zero-mode} LLs,\cite{w5,w12} under applied magnetic field $B$. Below a critical field value $B_c$, the lowest zero-mode LL has electron-like character and arises from the valence band, while the highest zero-mode LL has an heavy hole-like character and splits from the conduction band. In these inverted band conditions, the topological edge states are still present, although they are no longer protected by time-reversal symmetry.\cite{w32,w33,w33a} With increasing magnetic field, the zero-mode LLs cross each other at $B=B_c$. Above this magnetic field value the band structure becomes normal and only trivial quantum Hall insulator can be found.

The changing of external parameters may offer an effective way for fine tuning of phase transition between BI, TI and SM phases keeping intrinsic parameters of the QW. The latter can be highly desirable for future topological devices.\cite{w29} It has been recently demonstrated that a transition between BI and TI phases can be driven either by electric field\cite{w16,w17}, applied along the growth direction, or by temperature\cite{w17a,w17b}. However, the mentioned works were focused on the evolution of the band structure due to transition between BI and TI phases only, while temperature and electric field effects in the SM phase were ignored.

In this work, we are not only focused on temperature effects on the \emph{non-local} band structure but also propose to use hydrostatic pressure for fine tuning of the transitions between BI, TI and SM phases. We discover that at reasonable values of hydrostatic pressure and temperature, the band inversion at the $\Gamma$ point of the Brillouin zone does not lead to the formation of TI phase in HgTe QWs. We also show the evolution of the critical magnetic field $B_c$ with pressure and temperature in HgTe QWs of different width.



\onecolumngrid
\twocolumngrid

\section{\label{sec:Theory} Theoretical basis}
To describe the subband nonparabolicity\cite{w12,w13,w15,w21} and the spin-orbit interaction effects in HgTe QWs, we use 8-band \textbf{k$\cdot$p} Hamiltonian for the envelope wave functions, which takes into account the interaction between $\Gamma_6$, $\Gamma_8$ and $\Gamma_7$ bands. Further, we consider QWs grown on the [0$lk$] plane ($l$ and $k$ are integer numbers), which includes the most prevalent cases of (001)-\cite{w5,w6,w12,w21,w17b} and (013)-oriented\cite{w13,w15,w15a,w15b} structures. We note that previously reported 6-band \textbf{k$\cdot$p} Hamiltonian\cite{w37,w27a} for general (0$lk$) orientation considers only $\Gamma_6$ and $\Gamma_8$ bands and includes interaction with the $\Gamma_7$ band only via the second order perturbation theory.

Although electronic states in HgTe QWs can be indeed qualitatively described by the 6-band \textbf{k$\cdot$p} Hamiltonian, to calculate quantitative values of hydrostatic pressure and temperature, corresponding to the transition between BI, TI and SM phases, we also include the $\Gamma_7$ band in the Hamiltonian. The latter has significant effect on the electron-like states in HgTe QWs\cite{w34}, while for the band structure of the bulk HgCdTe-based materials, effect of the $\Gamma_7$ band can be neglected.\cite{w38}

The matrix elements of the 8-band \textbf{k$\cdot$p} Hamiltonian depends on momentum operators $k_x$, $k_y$, $k_z$, conduction and valence band edges $E_{c}(z)$ and $E_{v}(z)$, spin-orbit energy $\Delta(z)$, Kane energy $E_P$, and the modified Luttinger parameters $\gamma_1$, $\gamma_2$, $\gamma_3$, $\kappa$ and $F$. We also take into account effects of strain, resulting from mismatch of the lattice parameters in the given layer of heterostructure $a_L$ and the buffer $a_0$. The strain terms in the Hamiltonian include the hydrostatic $a_c$, $a_v$ and uniaxial $b$ and $d$ deformation potentials, as well as the strain tensor with components $\epsilon_{ij}$. In the work, we focus on HgTe/Cd$_{0.7}$Hg$_{0.3}$Te QWs grown on CdTe buffer. The explicit form of the 8-band \textbf{k$\cdot$p} Hamiltonian is given in the supplemental material.\cite{w34} We should point out that we have neglected in our Hamiltonian the linear-in-$\textbf{k}$ terms, resulting from the bulk inversion asymmetry (BIA) in bulk zinc-blende crystals.\cite{bk3} In the models with BIA\cite{w15b}, the crossing between zero-mode LLs at $B_c$ is avoided. The latter gives rise to specific behavior of magnetooptical transitions from the zero-mode LLs in the vicinity of critical magnetic field.\cite{w15,w15b,w14} If magnetic field either exceeds $B_c$ or remains significantly lower than the critical field, effects of BIA are negligible. Further, effects of BIA are therefore ignored.

To perform band structure calculations of HgTe/Cd(Hg)Te QWs at different values of hydrostatic pressure $P$ and temperature $T$, we take into account the dependencies of all relevant band parameters and the changes in the valence band offset (VBO) $\Omega$ between HgTe and CdTe. The VBO defines the misalignment of energy band gaps in adjacent layers of the QW. By setting the energy of $\Gamma_8$ band at $k=0$ in unstrained bulk HgTe to zero, $\Omega$ equals to the valence band edge $E_{v}(z)$ in the given layer.  According to Latussek \emph{et al.}\cite{w27} and Becker \emph{et al.},\cite{w28a} $\Omega$ has a linear dependence on $P$ and $T$:
\begin{equation}
\label{eq:5}
\Omega(P,T)=\Omega_0+\beta_{P}P+\beta_{T}T,
\end{equation}
where $\Omega_0$ is the VBO at $T=0$ and $P=0$ (counted from atmospheric pressure), while $\beta_{P}$ and $\beta_{T}$ are independent on pressure and temperature. It is worth noting that that $\beta_{P}$ is known only for VBO between HgTe and Cd$_{0.7}$Hg$_{0.3}$Te.\cite{w27} Thus, by assuming $\beta_{P}$ to vary linearly with $x$ for Cd$_x$Hg$_{1-x}$Te alloy, we extract the value for VBO between HgTe and CdTe.

\begin{table}
\caption{\label{tab:1}Band parameters for HgTe and CdTe independent of hydrostatic pressure and temperature.}
\begin{ruledtabular}
\begin{tabular}{cccccc}
Parameters & CdTe & HgTe & Parameters & CdTe & HgTe\\
\hline
$\alpha_0$ (meV/GPa) & 81\footnote[1]{Ref.~\onlinecite{w27}} & 87.2\footnotemark[1] & $\alpha_1$ (meV/GPa$^2$) & -4.96\footnotemark[1] & -4.61\footnotemark[1]\\
$\beta_T$ (meV/K) &0.4\footnote[2]{Ref.~\onlinecite{w28a}}& 0 &$\beta_P$ (meV/GPa) &35.7\footnote[3]{Calculated by using results of Ref.~\onlinecite{w27}}& 0\\
$\Omega_0$ (eV)&-0.57&0&$a_{300K}$ ({\AA})&6.4823\footnote[4]{Ref.~\onlinecite{bk1}}&6.4615\footnotemark[4] \\
$\Delta$ (eV)&0.91&1.08&$F$&0&-0.09\\
$E_P$ (eV)&18.8\footnote[5]{Recent results\cite{w30} show temperature independence of $E_P$ in HgCdTe alloys}&18.8\footnotemark[5]&$\kappa$&-1.31&-0.4\\
$dc_{11}/dP$&4.44\footnote[6]{Calculated by using results of Ref.~\onlinecite{w28c}}&3.30\footnote[7]{Ref.~\onlinecite{bk2}}&$\gamma_1$&1.47&4.1\\
$dc_{12}/dP$&2.62\footnotemark[6]&4.10\footnotemark[7]&$\gamma_2$&-0.28&0.5\\
$dc_{44}/dP$&1.92\footnotemark[6]&-0.12\footnotemark[7]&$\gamma_3$&0.03&1.3\\
$a_c$ (eV)&-2.925&-2.380&$a_v$ (eV)&0&1.31\\
$b$ (eV)&-1.2&-1.5&$d$ (eV)&-5.4&-2.5\\
\end{tabular}
\end{ruledtabular}
\end{table}

Normally the pressure dependence of the band gap $E_g=E_{c}-E_{v}$ in bulk materials is analyzed by means of a quadratic equation:
\begin{equation}
\label{eq:6}
E_{g}(P,T)=E_{g}^{(0)}(T)+\alpha_{0}P+\alpha_{1}P^2,
\end{equation}
in which $\alpha_{0}$ and $\alpha_{1}$ depend on the range of pressures over which the analysis is conducted.\cite{w27} Previously, $\alpha_{0}$ and $\alpha_{1}$ for bulk HgTe have been determined from pressure dependence of intersubband transitions in HgTe/Cd$_{0.7}$Hg$_{0.3}$Te superlattices at $P<2.5$ GPa.\cite{w27} By analyzing experimental results, $\alpha_{0}$ and $\alpha_{1}$ were obtained to be equal to 87.2 meV/GPa and -4.61 meV/GPa$^2$ respectively. For bulk CdTe, in accordance with Eq.~(7) and (8) in Ref.~\onlinecite{w27}, we set $\alpha_{0}$ and $\alpha_{1}$ meV/GPa and -4.96 meV/GPa$^2$. The temperature dependence of $E_{g}^{(0)}(T)$ on $x$ for Cd$_x$Hg$_{1-x}$Te alloy is determined from the empirical expression according to Laurenti \emph{et al.}\cite{w28}

Previous studies have shown pronounced dependence of overlapping between valence and conduction bands on lattice-mismatch deformation in the SM phase in wide HgTe QW.\cite{w13a} Therefore, to describe the temperature and pressure effects on electronic states in the SM phase, one should accurately take into account the dependence of elastic constants and lattice parameters on $P$ and $T$. In this paper, the elastic constants $c_{ij}$ are assumed to have linear dependence on $P$\cite{w28c,bk2}:
\begin{equation}
\label{eq:7}
c_{ij}(P,T)=c_{ij}^{(0)}(T)+P\dfrac{dc_{ij}}{dP},
\end{equation}
while the temperature dependencies of $c_{ij}^{(0)}(T)$ describe experimental data.\cite{bk1} Here, $dc_{ij}/dP$ are assumed to be independent on $P$ and $T$. The linear dependence on $P$ in Eq.~\eqref{eq:7} allows us to use Murnaghan's equation of state for changing in the lattice parameter $a_L$ in the given layer with hydrostatic pressure:
\begin{equation}
\label{eq:8}
a_{L}(P,T)=a_{L}^{(0)}(T)\left[1+P\dfrac{B'_0}{B_0(T)}\right]^{-1/3B'_0},
\end{equation}
where $B_0(T)=(c_{11}(T)+2c_{12}(T))/3$ is the bulk modulus and $B'_0=(dc_{11}/dP+2dc_{12}/dP)/3$.

Temperature dependence of $a_{L}^{(0)}(T)$ is obtained by solving the differential equation
\begin{equation}
\label{eq:9}
\alpha(T)=\dfrac{1}{a_{L}^{(0)}(T)}\dfrac{da_{L}^{(0)}(T)}{dT}
\end{equation}
with condition
\begin{equation*}
a_{L}^{(0)}(300K)=a_{300K}.
\end{equation*}
Here $\alpha(T)$ is the thermal expansion coefficient, taken from experimental data.\cite{bk1} Other band structure parameters for HgTe and CdTe, which are supposed to be independent on $T$ and $P$, are listed in Table~\ref{tab:1}. As in Ref.~\onlinecite{w21}, the band parameters are assumed to be a piecewise function along the growth direction and to vary linearly with $x$ in Cd$_x$Hg$_{1-x}$Te alloy.

\begin{figure}
\includegraphics [width=1.0\columnwidth, keepaspectratio] {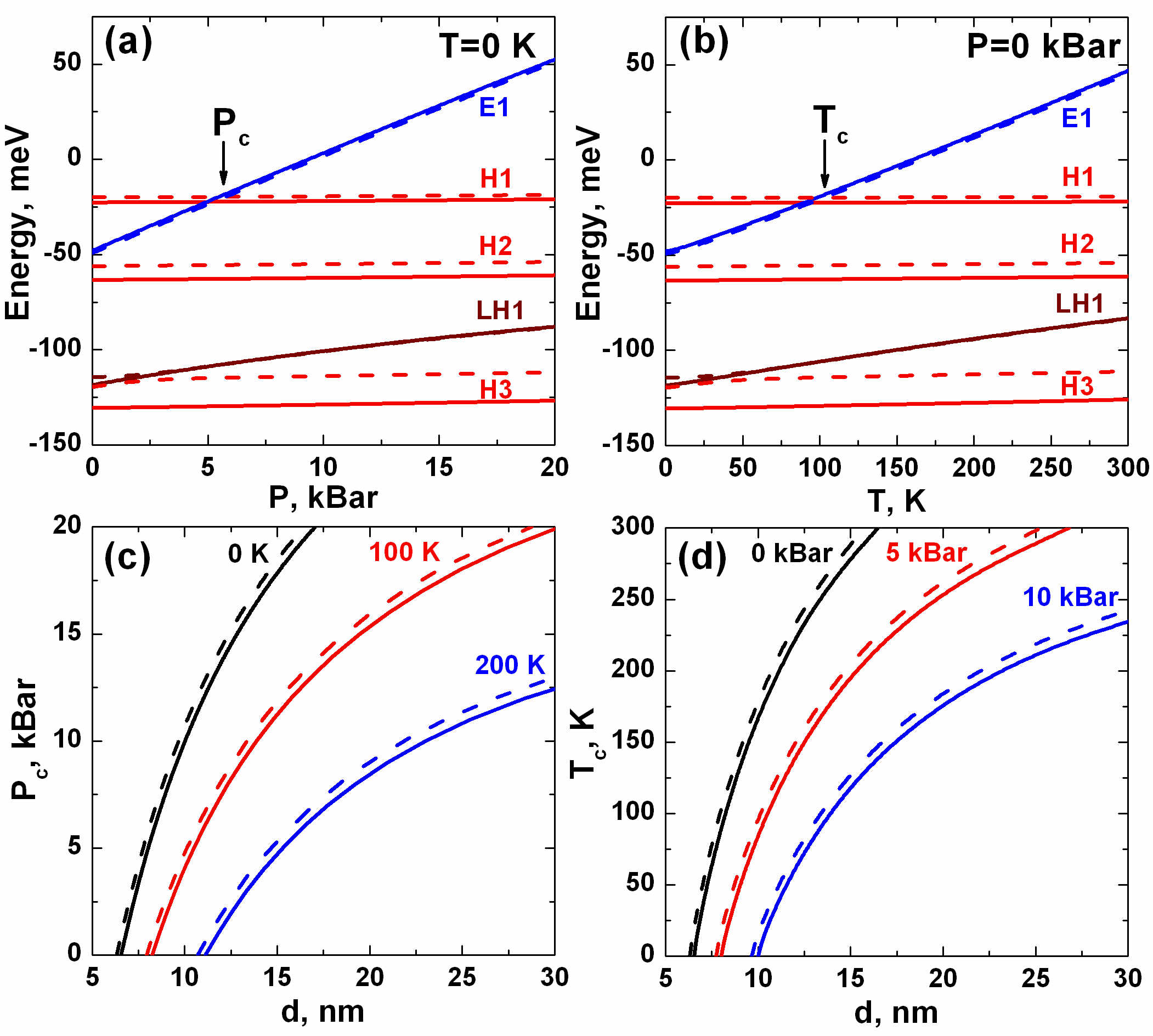} 
\caption{\label{Fig:2} (Color online) (a,b) Band edge of the electron-like E1, the heavy-hole-like H1, H2, and H3 and the light-hole-like LH1 subbands at the $\Gamma$ point in 8 nm HgTe/Cd$_{0.7}$Hg$_{0.3}$Te QWs as a function of hydrostatic pressure (a) and temperature (b). (c,d) Dependence of critical pressure $P_c$ and temperature $T_c$, at which the Dirac cone in the $\Gamma$ point arises (see Fig.~\ref{Fig:1}b), as a function of $d$ for HgTe/Cd$_{0.7}$Hg$_{0.3}$Te QWs. In all the panels, solid curves represent the calculations for (001)-oriented QWs, while the dashed curves correspond to the structures grown on [013] plane.}
\end{figure}

\section{\label{sec:RnD}Results and discussions}
\subsection{Topological insulator phase}
Inversion of electronic subbands at the $\Gamma$ point and non-zero band gap are essential for a formation of TI phase in HgTe QWs.\cite{w4,w5} Therefore, we first focus on the effect of changing of the subband ordering with hydrostatic pressure and temperature. The top panels in Fig.~\ref{Fig:2} present energies of the electron-like E1, the heavy-hole-like H1, H2, H3 and the light-hole-like LH1 subbands at the $\Gamma$ point as a function of $P$ and $T$, calculated for 8 nm HgTe/Cd$_{0.7}$Hg$_{0.3}$Te QWs. This figure shows the similarity of pressure and temperature effects on the band ordering.

At small values of $P$ and $T$, the band structure remains inverted and TI phase in HgTe QW survives. However, strong pressure and temperature dependence of the E1 subband results in the crossing between E1 and H1 subbands at some critical values of pressure $P_c$ and temperature $T_c$. In this case, energy dispersion in the vicinity of the $\Gamma$ point is linear in quasimomentum (see Fig.~\ref{Fig:1}b), and further increasing of $P$ and $T$ puts the HgTe QW into BI phase with direct band ordering. The difference in $P_c$ and $T_c$ for (001) and (013)-oriented QWs is mostly related with lattice-mismatch strain, which also depends on the growth direction (see Eq.~(3) in Ref.\cite{w34}). Two bottom panels in Fig.~\ref{Fig:2} demonstrate that both $P_c$ and $T_c$ has a strong nonlinear dependence on the QW width. Moreover, $P_c$ and $T_c$ depend on temperature and pressure respectively; both quantities significantly decrease with $T$ and $P$ at given QW width.


\begin{figure}
\includegraphics [width=1.0\columnwidth, keepaspectratio] {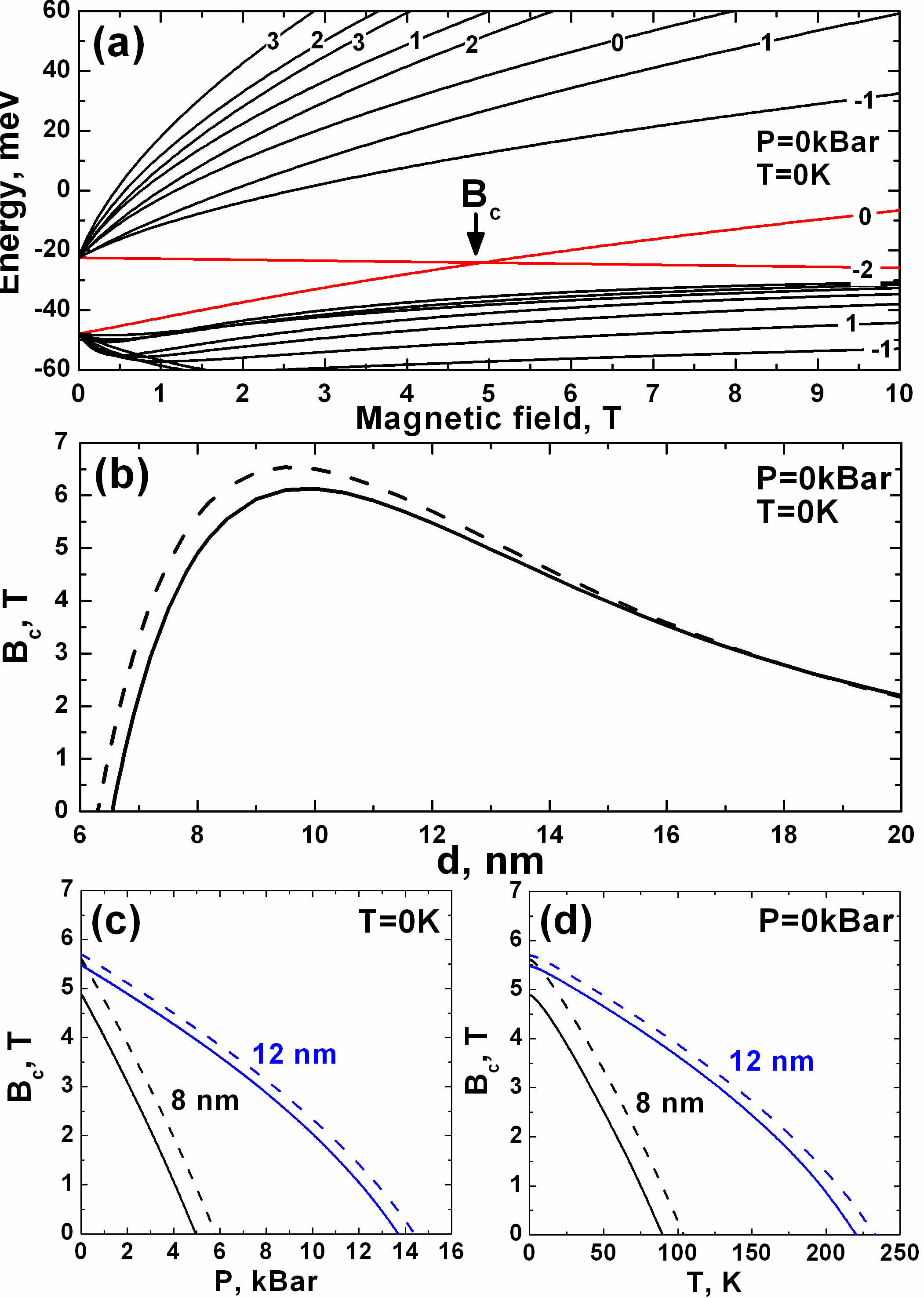} 
\caption{\label{Fig:5} (Color online) (a) Landau levels for (001)-oriented HgTe/Cd$_{0.7}$Hg$_{0.3}$Te QW of 8 nm thickness at $T=$0 K and $P=$0 kBar. Pair of zero-mode LLs is shown by red curves. (b) A critical field $B_c$ as a function of QW width $d$ at zero pressure and temperature. (c,d) A critical field $B_c$ as a function of hydrostatic pressure and temperature for 8 nm and 12 nm QWs. The solid curves in the panels (b-d) display the calculations for (001)-oriented QWs, while the dashed curves correspond to the (013)-oriented structures.}
\end{figure}

As mentioned above, the band inversion at the $\Gamma$ point leads to the crossing of zero-mode LLs in critical magnetic field $B_c$.\cite{w5,w12} If $B>B_c$, HgTe QW has a direct band ordering, while below $B_c$ the band structure remains inverted and the helical edge states still exists.\cite{w32,w33,w33a} The top panel in Fig.~\ref{Fig:5} shows LLs in (001) HgTe/Cd$_{0.7}$Hg$_{0.3}$Te QW calculated for zero $P$ and $T$ values. The numbers over the curves correspond to the LL indices.\cite{w34} Two red curves are the zero-mode LLs, which are identified within a simplified approach, based on 2D Dirac-type Hamiltonian.\cite{w4} The panel below displays $B_c$ as a function of QW width $d$ for the (001)- and (013)-oriented QWs for zero pressure and temperature.

The bottom panels in Fig.~\ref{Fig:5} show the dependence of $B_c$ on $P$ and $T$ for the 8 nm and 12 nm QWs of different orientations. It is seen that critical magnetic field decreases with pressure and temperature. The latter is related with the collapse of the gap between E1 and H1 subbands, if pressure and temperature tends to $P_c$ and $T_c$ respectively (see Fig.~\ref{Fig:2}). Above $P_c$ and $T_c$, the band structure is direct and the zero-mode LLs are not crossed.\cite{w12}

So far, we have considered ordering of electronic subbands in HgTe QWs in the vicinity of $k=0$, driven by hydrostatic pressure and temperature. However, such \emph{local} picture does not account all the electronic properties of HgTe QWs in the TI phase. For instance, if the width $d$ increases, the QW has indirect band gap due to arising of the side maxima in the valence band, whose positions depend on the growth direction.\cite{w39}

\begin{center}
\begin{figure}[h]
  \begin{minipage}{0.497\linewidth} 
\center{\Large{\textbf{(a)}} \includegraphics [width=1.0\columnwidth, keepaspectratio] {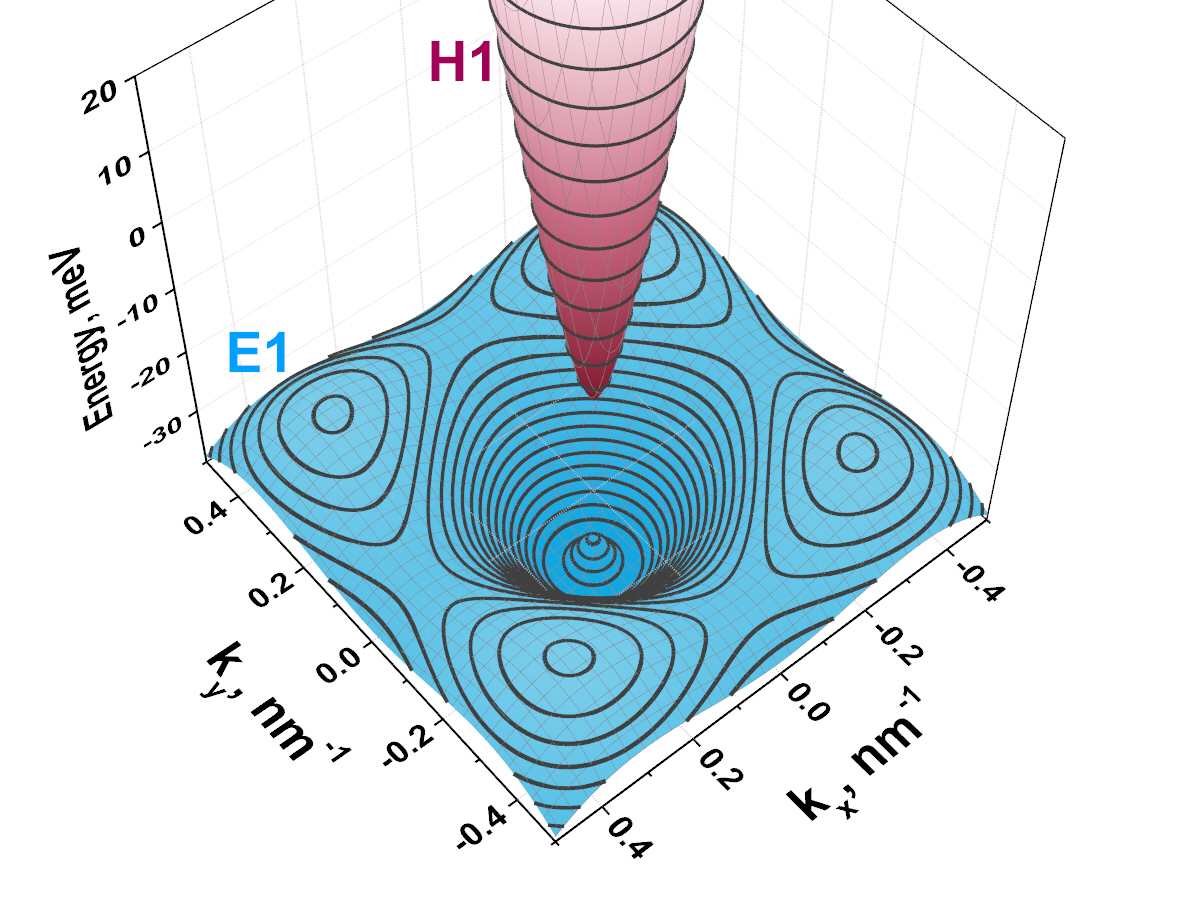}} 
   \end{minipage}
    \hfill
  \begin{minipage}{0.497\linewidth} 
\center{\Large{\textbf{(b)}} \includegraphics [width=1.0\columnwidth, keepaspectratio] {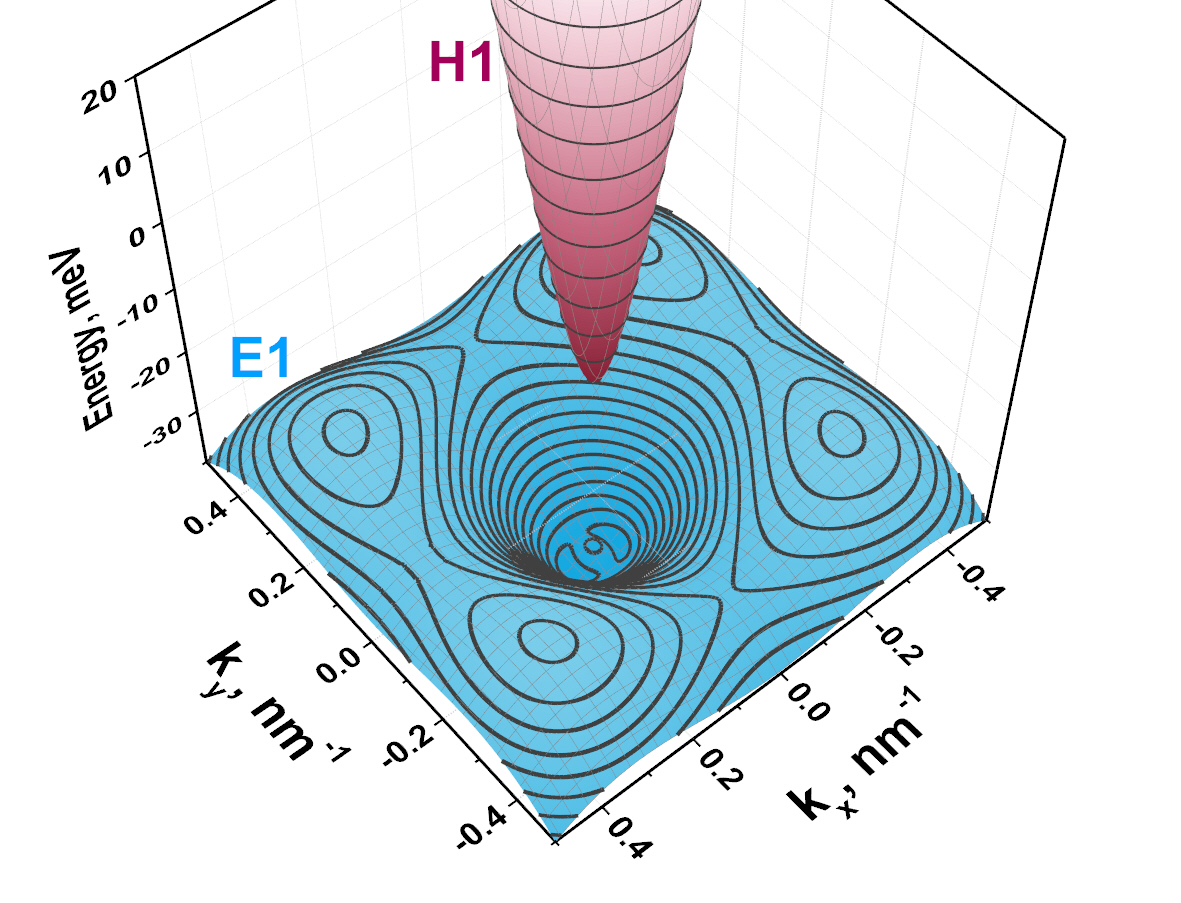}} 
   \end{minipage}
\caption{\label{Fig:3} (Color online) Band structure and contour lines for HgTe/Cd$_{0.7}$Hg$_{0.3}$Te QW of 8 nm width, grown on (a) [001] and (b) [013] planes for $T$=0 and $P$=0. The E1 subband is shown in blue, the red surface corresponds to the H1 subband. The $x$ and $y$ axes for the (001) QW are oriented along (100) and (010) crystallographic directions, while for the (013) QW, the axes correspond to the directions of (100) and (0$3\bar{1}$), respectively.}
\end{figure}
\end{center}

To illustrate the differences, arising in valence bands for (001)- and (013)-oriented HgTe QWs, we provide a 3D plot of the band structure and contour lines for the 8 nm QWs at $P=0$ and $T=0$ (see Fig.~\ref{Fig:3}). For both QWs, conduction band has an isotropic energy-momentum law. The valence band is anisotropic with four side maxima shifted from the $\Gamma$ point. However, the valence band in the (013)-oriented QW is highly anisotropic even at small values of quasimomentum. Due to low-symmetry growth orientation, positions of the four side maxima in (013)-oriented QWs depend also on the QW width. Thus, to calculate the values of indirect band gap in (013)-oriented QW, one should first find the crystallographic direction, corresponding to the side maxima at given QW width.

\onecolumngrid
\twocolumngrid

\begin{figure}
\includegraphics [width=1.0\columnwidth, keepaspectratio] {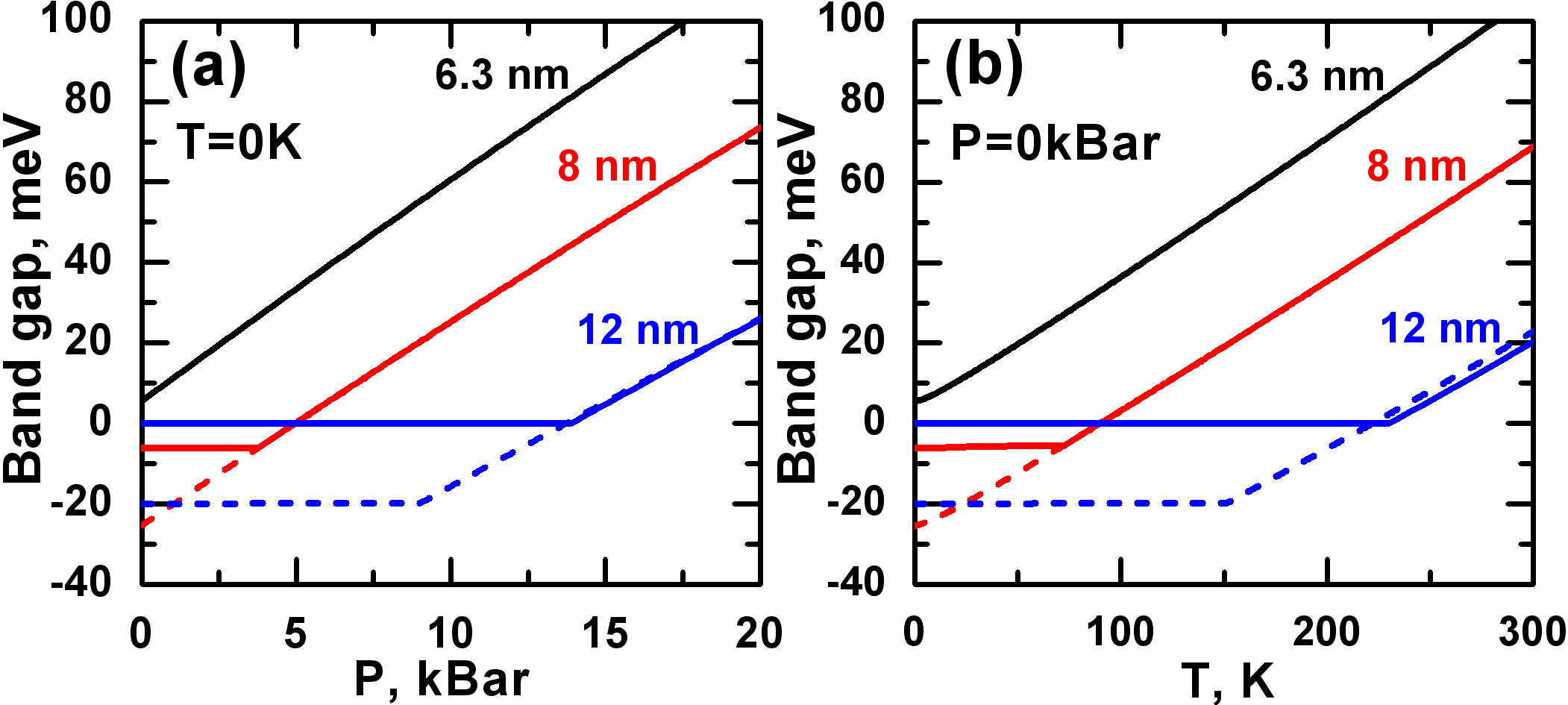} 
\caption{\label{Fig:4} (Color online) The band gap in HgTe/Cd$_{0.7}$Hg$_{0.3}$Te QWs grown on (001) CdTe buffer as a function of (a) $P$ and (b) $T$, calculated for different QW widths. The negative band gap values correspond to inverted band structure. The dashed curves is evolution of the gap at the $\Gamma$ point.}
\end{figure}

On the other hand, positions of the maxima in the (001) QWs are independent of the QW width. They always lie along (110), ($\bar{1}$10), (1$\bar{1}$0), ($\bar{1}\bar{1}$0) directions, which are all equivalent. The latter is caused by the fact that anisotropic terms  in the 8-band \textbf{k$\cdot$p} Hamiltonian for (001)-oriented structures are independent of $k_z$\cite{w21}. Because of the easier calculations, we consider the pressure and temperature effects on the \emph{non-local} band structure only for (001)-oriented HgTe QWs. Such effects for (013)-oriented QWs are expected to be qualitatively the same. We want to stress that all the disparities between different orientations are related with the valence band. They concern two main characteristics: i) positions of the side maxima in the valence band and ii) anisotropy of energy dispersion in the valence subband at small quasimomentum. However, detailed quantitative calculations for such low-symmetry orientation are very time-consuming.

Fig.~\ref{Fig:4} shows indirect band gap in HgTe QWs of different widths as a function of hydrostatic pressure and temperature. It is clear that the band gap evolution strongly depends on the QW width. For the QWs with direct band gap, the band gap increases with $P$ and $T$: from the negative values if the band structure is inverted, and from the positive values in the case of the band insulator phase. If HgTe QW is wide enough, the system has indirect band gap, which has a weaker dependence on $P$ and $T$ than the gap at the $\Gamma$ point (see the 8 nm thick QW). At specific values of pressure and temperature, the side maxima in the valence band are placed below the top at $k=0$. Further increasing of $P$ or $T$ results in rising of the band gap almost linearly with pressure and temperature.

The indirect gap in the 12 nm HgTe QW equals to zero in a wide range of pressure and temperature. This holds as long as the gap at the $\Gamma$ point does not vanishe (see the dashed curves in Fig.~\ref{Fig:4}). Further increasing of $P$ and $T$ puts the system into the BI phase with a sub-linear dependence of the band gap on hydrostatic pressure and temperature. We note that the gap at the $\Gamma$ point also features a non-monotonic behavior, which is related to swapping of E1 and H2 subbands. The indirect band gap is very sensitive to the strain effects resulted from difference in lattice constants in the QW, barriers and buffer. For instance, a 12 nm HgTe/Cd$_{0.7}$Hg$_{0.3}$Te QW grown on CdTe buffer has zero indirect band gap at small values of $P$ and $T$ (see Fig.~\ref{Fig:4}), while for the QW grown on another buffer, the band gap may be opened.\cite{w17b}

\begin{figure}
\includegraphics [width=1.0\columnwidth, keepaspectratio] {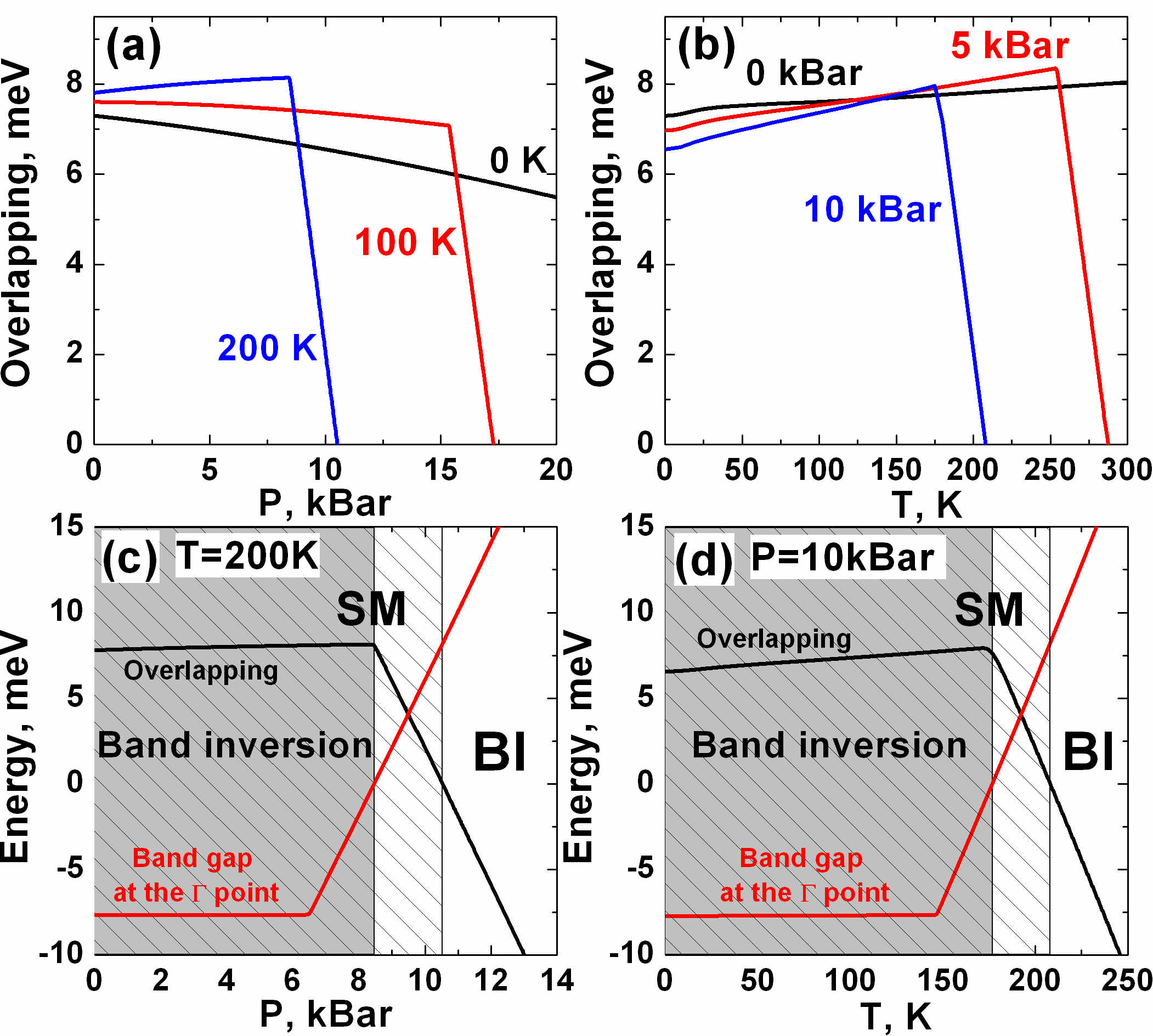} 
\caption{\label{Fig:6} (Color online) (a,b) Overlapping between conduction and valence subbands in 20 nm HgTe/Cd$_{0.7}$Hg$_{0.3}$Te QW, grown on (001) CdTe buffer as a function of (a) hydrostatic pressure and (b) temperature. (c,d) Overlapping (black curves) and band gap at the $\Gamma$ point (red curves): (c) as a function of $P$ for $T=200$ K and (d) as a function of $T$ for $P=10$ kBar. The negative overlapping values in the panels (c,d) define indirect band gap. The white-open regions are connected with the BI phase, the striped regions correspond to the SM phase, while the grey regions define the range of $P$ and $T$ with the inverted band structure.}
\end{figure}


\subsection{Semimetal phase}

In wide HgTe QWs indirect band gap vanishes and the system is characterized by non-local overlapping between conduction and valence bands. As mentioned above, the latter case is called the SM phase.\cite{w13,w13a} Typically, it arises when the E1 level lies below several heavy-hole-like subbands. Let us now focus on the pressure and temperature evolution of the overlapping between conduction and valence bands in the SM phase.

For this purpose, we consider a 20 nm HgTe QW, in which existence of SM phase at low temperatures was demonstrated in number of experiments.\cite{w13,w13a,w13b,w13c,w13d} Fig.~\ref{Fig:6} represents the calculations of non-local overlapping between conduction and valence subbands as a function of $P$ and $T$. In contrast to the band gap evolution, pressure and temperature have different effects on the band overlapping. Increasing of the pressure reduces the overlapping at $T=0$, while the temperature increases the overlapping values for zero hydrostatic pressure. Variation of both $P$ and $T$ may modify the evolution of overlapping significantly. The latter is demonstrated by the red and blue curves in Fig.~\ref{Fig:6}. It is seen that at specific values of $P$ and $T$, the band overlapping decreases dramatically.


\onecolumngrid
\begin{center}
\begin{figure}[h]
\includegraphics [width=1.0\columnwidth, keepaspectratio] {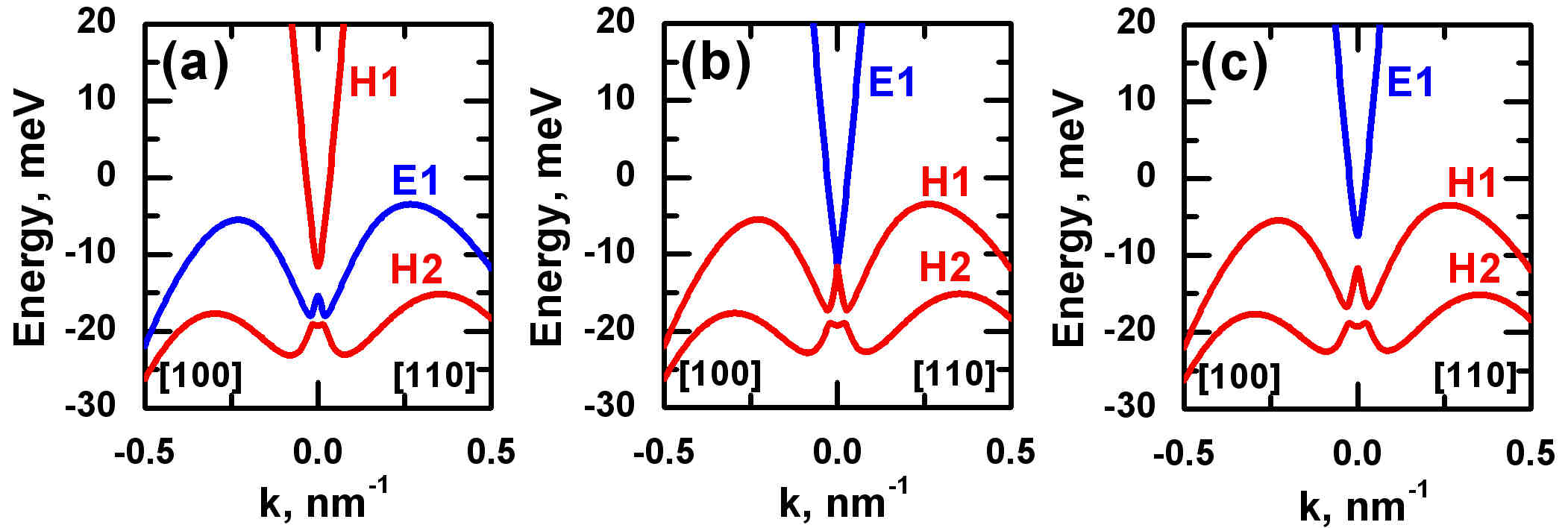} 
\caption{\label{Fig:7} (Color online) Evolution of the band structure in 20 nm  HgTe/Cd$_{0.7}$Hg$_{0.3}$Te QW, grown on (001) CdTe buffer, at $T=200$~K with hydrostatic pressure: (a) $P=7.5$~kBar (the SM phase with inverted band structure); (a) $P=8.45$~kBar (the SM phase with the Dirac cone in the $\Gamma$ point); (c) $P=9.5$~kBar (the SM phase with direct band ordering). Electron-like and heavy-hole-like subbands are shown in blue and red respectively.}
\end{figure}
\begin{figure}[h]
\includegraphics [width=1.0\columnwidth, keepaspectratio] {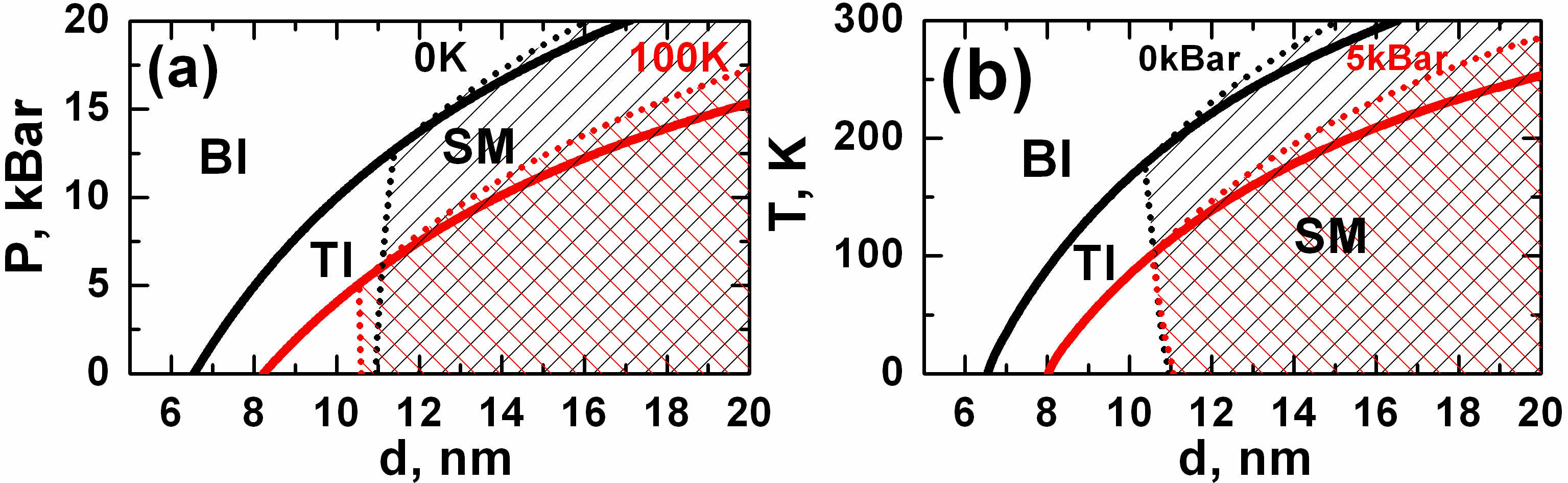} 
\caption{\label{Fig:8} (Color online) (a) Pressure and (b) temperature phase diagram for (001) HgTe/Cd$_{0.7}$Hg$_{0.3}$Te QWs, grown on CdTe buffer. The solid curves correspond to the arising of the Dirac cone at the $\Gamma$ point. The dotted curves conform to a formation of the gapless states, shown in Fig.~\ref{Fig:1}d.}
\end{figure}
\end{center}
\twocolumngrid
The origin of such decrease is related with a change of the band structure from inverted into the normal one under joint effect of pressure and temperature. Two bottom panels in Fig.~\ref{Fig:6} show pressure and temperature evolution of the overlapping and the band gap at the $\Gamma$ point at $T=200$~K and $P=10$~kBar respectively. One can see that, at the values of $P$ and $T$, corresponding to the decreasing of the band overlapping, the band gap energy at the $\Gamma$ point changes sign. We note that the normal band ordering is defined by positive band gap at the $\Gamma$ point even if the indirect band gap equals to zero. Thus, the white-striped regions in Fig.~\ref{Fig:6}c and \ref{Fig:6}d, characterized by both positive values of the overlapping and the gap at $k=0$, correspond to the SM phase but with \emph{normal} band ordering.

The SM phase with direct band ordering, arising under hydrostatic pressure and temperature, has not been predicted before. Evolution of the band structure of 20 nm HgTe QW due to pressure-driven phase transition into this specific phase at $T=200$~K is shown in Fig.~\ref{Fig:7}. It is seen that the band inversion at the $\Gamma$ point is accompanied by non-local overlapping of the conduction and valence bands. We note that such phase transition can not be described within the simplified 2D Dirac-type model,\cite{w4} because it considers the bulk and edge states only in the vicinity of $k=0$. Moreover, any known simplified 2D models\cite{w4,w35,w36,w37} are not generally applicable to this case. Therefore, even qualitative picture of the edge states in the SM phase is unknown.

Fig.~\ref{Fig:8} presents pressure and temperature phase diagrams for (001) HgTe/Cd$_{0.7}$Hg$_{0.3}$Te QWs grown on CdTe buffer. The white-open regions in all the panels correspond to the insulator phases, while the striped regions are the SM phase with the overlapping of conduction and valence bands. First, it is seen that in a wide range of QW width $d$, tuning of hydrostatic pressure and temperature allows one to drive transitions between SM, BI and TI phases. The second point is that TI phase in HgTe QWs exists only in the finite ranges of $P$ and $T$. Indeed, at given value of hydrostatic pressure, there is a limit temperature above which the TI phase collapses due to non-local overlapping between the conduction and valence bands. Our calculations, performed for HgTe/Cd$_{x}$Hg$_{1-x}$Te QWs at different values of Cd concentration $x$, show that such limit temperature value does not exceed room temperature. The diagrams in Fig.~\ref{Fig:8} evidence that variation of $d$ at high pressure and temperature drives the system from BI into SM, avoiding a TI phase.

\section{\label{sec:Summary}Conclusions}
With accurate calculations on the basis of the 8-band \textbf{k$\cdot$p} Hamiltonian, we have theoretically studied effect of hydrostatic pressure and temperature on the band structure and Landau levels in HgTe QWs, grown along (001) and (013) crystallographic orientations. We have demonstrated that variation of these two external parameters can be efficiently used for driving transitions between semimetal, band insulator and topological insulator phases. We have shown the existence of topological insulator phase only in the finite range of $P$ and $T$. At high pressure and temperature, variation of HgTe QW width drives the system from band insulator into the semimetal, avoiding the topological insulator phase.

At specific values of pressure and temperature, our band structure calculations reveal that the band inversion in HgTe QWs does not lead to formation of topological insulator phase due to accompanying non-local overlapping between conduction and valence bands. The pressure and temperature diagrams for (001) HgTe/Cd$_{0.7}$Hg$_{0.3}$Te QWs grown on CdTe buffer have been presented. Our results provide a theoretical basis for future magnetotransport and magnetospectroscopy experimental works.\cite{w40,w41}

\begin{acknowledgments}
This work was supported by the CNRS through LIA TeraMIR project and the post-doc prolongation program of the Institute of Physics, by the Languedoc-Roussillon region via the Gepeto Terahertz platform, the Russian Academy of Sciences, the non-profit Dynasty foundation, the Russian Foundation for Basic Research (Grant Nos. 15-02-08274, 15-52-16012 and 16-02-00672), the Russian Scientific Foundation (Grant No 16-12-10317), by Russian Ministry of Education and Science (Grant Nos. MK-6830.2015.2, HIII-1214.2014.2) and by the HARMONIA project of National Science Center (Poland), allocated on the basis of decision number DEC-2013/10/M/ST3/00705.
\end{acknowledgments}


%
\clearpage
\onecolumngrid
\section*{Supplemental Material}
\maketitle
\onecolumngrid
\subsection{The 8-band \textbf{k$\cdot$p} Hamiltonian}
This supplemental material presents the general form of the 8-band \textbf{k$\cdot$p} Hamiltonian for strained (0$lk$) heterostructures ($l$, $k$ are integer) and details of band structure and Landau levels calculations. Neglecting the small terms, resulting from the lack of inversion symmetry in bulk zinc-blende crystals,\cite{bk3} in given basis\cite{w21} of the Bloch amplitudes for the $\Gamma_6$, $\Gamma_8$ and $\Gamma_7$ bands
\begin{equation*}
  U_1(\textbf{r})=|\Gamma_6,+1/2\rangle = S\uparrow;
\end{equation*}
\begin{equation*}
  U_2(\textbf{r})=|\Gamma_6,-1/2\rangle = S\downarrow;
\end{equation*}
\begin{equation*}
  U_3(\textbf{r})=|\Gamma_8,+3/2\rangle = \frac{1}{\sqrt{2}}(X+iY)\uparrow;
\end{equation*}
\begin{equation*}
  U_4(\textbf{r})=|\Gamma_8,+1/2\rangle = \frac{1}{\sqrt{6}}[(X+iY)\downarrow - 2Z\uparrow];
\end{equation*}
\begin{equation*}
  U_5(\textbf{r})=|\Gamma_8,-1/2\rangle = -\frac{1}{\sqrt{6}}[(X-iY)\uparrow + 2Z\downarrow];
\end{equation*}
\begin{equation*}
  U_6(\textbf{r})=|\Gamma_8,-3/2\rangle = -\frac{1}{\sqrt{2}}(X-iY)\downarrow;
\end{equation*}
\begin{equation*}
  U_7(\textbf{r})=|\Gamma_7,+1/2\rangle = \frac{1}{\sqrt{3}}[(X+iY)\downarrow + Z\uparrow];
\end{equation*}
\begin{equation*}
  U_8(\textbf{r})=|\Gamma_7,-1/2\rangle = \frac{1}{\sqrt{3}}[(X-iY)\uparrow - Z\downarrow];
\end{equation*}
the 8-band \textbf{k$\cdot$p} Hamiltonian takes the form
\begin{equation}\label{eq:1}
  \hat{H}_{\textbf{k}}^{8\times 8} = \begin{pmatrix}
    \hat{T} & 0 &
    -\frac{1}{\sqrt{2}}\tilde{P}k_+ & \sqrt{\frac{2}{3}}\tilde{P}\hat{k}_z &
    \frac{1}{\sqrt{6}}\tilde{P}k_- & 0 &
    -\frac{1}{\sqrt{3}}\tilde{P}\hat{k}_z & -\frac{1}{\sqrt{3}}\tilde{P}k_-
    \\
    0 & \hat{T} &
    0 & -\frac{1}{\sqrt{6}}\tilde{P}k_+ &
    \sqrt{\frac{2}{3}}\tilde{P}\hat{k}_z & \frac{1}{\sqrt{2}}\tilde{P}k_- &
    -\frac{1}{\sqrt{3}}\tilde{P}k_+ & \frac{1}{\sqrt{3}}\tilde{P}\hat{k}_z
    \\
    -\frac{1}{\sqrt{2}}\tilde{P}k_- & 0 &
    \hat{U}+\hat{V} & -\bar{S}_- & \hat{R} & 0 &
    \frac{1}{\sqrt{2}}\bar{S}_- & -\sqrt{2}\hat{R}
    \\
    \sqrt{\frac{2}{3}}\tilde{P}\hat{k}_z & -\frac{1}{\sqrt{6}}\tilde{P}k_- &
    -\bar{S}_-^\dagger & \hat{U}-\hat{V} & \hat{C} & \hat{R} &
    \sqrt{2}\hat{V} & -\sqrt{\frac{3}{2}}\tilde{S}_-
    \\
    \frac{1}{\sqrt{6}}\tilde{P}k_+ & \sqrt{\frac{2}{3}}\tilde{P}\hat{k}_z &
    \hat{R}^\dagger & \hat{C}^\dagger & \hat{U}-\hat{V} & \bar{S}_+^\dagger &
    -\sqrt{\frac{3}{2}}\tilde{S}_+ & -\sqrt{2}\hat{V}
    \\
    0 & \frac{1}{\sqrt{2}}\tilde{P}k_+ &
    0 & \hat{R}^\dagger & \bar{S}_+ & \hat{U}+\hat{V} &
    \sqrt{2}\hat{R}^\dagger & \frac{1}{\sqrt{2}}\bar{S}_+
    \\
    -\frac{1}{\sqrt{3}}\tilde{P}\hat{k}_z & -\frac{1}{\sqrt{3}}\tilde{P}k_- &
    \frac{1}{\sqrt{2}}\bar{S}_-^\dagger & \sqrt{2}\hat{V} &
    -\sqrt{\frac{3}{2}}\tilde{S}_+^\dagger & \sqrt{2}\hat{R} &
    \hat{U}-\Delta & \hat{C}
    \\
    -\frac{1}{\sqrt{3}}\tilde{P}k_+ & \frac{1}{\sqrt{3}}\tilde{P}\hat{k}_z &
    -\sqrt{2}\hat{R}^\dagger & -\sqrt{\frac{3}{2}}\tilde{S}_-^\dagger &
    -\sqrt{2}\hat{V} & \frac{1}{\sqrt{2}}\bar{S}_+^\dagger &
    \hat{C}^\dagger & \hat{U}-\Delta
  \end{pmatrix}
\end{equation}
where $\varphi=\arctan(l/k)$ is the angle between the growth direction (0$lk$) and the (001) axis,
\begin{equation*}
  \hat{T} = E_c + \frac{\hbar^2}{2m_0}\left\{
    (2F+1)(k_x^2 + k_y^2)
    + \hat{k}_z (2F+1) \hat{k}_z\right\}+\hat{T}^{(\epsilon)},
\end{equation*}
\begin{equation*}
  \hat{U} = E_v - \frac{\hbar^2}{2m_0}\left\{
    \gamma_1(k_x^2 + k_y^2)
    + \hat{k}_z \gamma_1 \hat{k}_z\right\}+\hat{U}^{(\epsilon)},
\end{equation*}
\begin{multline*}
  \hat{V} = -\frac{\hbar^2}{2m_0}\Biggl\{
    \gamma_2 (k_x^2 + k_y^2)-2\hat{k}_{z}\gamma_{2}\hat{k}_{z}-\frac{3}{2}\sin^2(2\varphi)\left(\gamma_2-\gamma_3\right) k_y^2+\\
    + \frac{3}{2}\sin^2(2\varphi)\hat{k}_z \left(\gamma_2-\gamma_3\right) \hat{k}_z
    + \frac{3}{4}\sin(4\varphi) k_y \{\gamma_2-\gamma_3,\hat{k}_z\}
  \Biggr\}+\hat{V}^{(\epsilon)},
\end{multline*}
\begin{multline*}
  \hat{R} = -\frac{\hbar^2}{2m_0}\sqrt{3}\Biggl\{
    \gamma_2(k_y^2-k_x^2)+2i\gamma_3 k_x k_y-\frac{1}{2}\sin^2(2\varphi)\left(\gamma_2-\gamma_3\right) k_y^2
    + \\
    + \frac{1}{2}\sin^2(2\varphi)\hat{k}_z(\gamma_2-\gamma_3) \hat{k}_z
    + \frac{1}{4}\sin(4\varphi) k_y \{\gamma_2-\gamma_3,\hat{k}_z\}
  \Biggr\}+\hat{R}^{(\epsilon)},
\end{multline*}
\begin{multline*}
  \bar{S}_\pm = -\frac{\hbar^2}{2m_0}\sqrt{3}\Biggl\{
    k_{\pm}\{\gamma_3,\hat{k}_z\}+k_{\pm}[\kappa,\hat{k}_z]\pm i k_y\sin^2(2\varphi) \Bigl\{\gamma_2-\gamma_3,\hat{k}_z\Bigr\}
    \pm \frac{i}{2}\sin(4\varphi)(\gamma_2-\gamma_3) k_y^2 \\
    \mp\frac{i}{2}\sin(4\varphi)\hat{k}_{z}(\gamma_2-\gamma_3)\hat{k}_z
  \Biggr\}+\hat{S}_{\pm}^{(\epsilon)},
\end{multline*}
\begin{equation}\label{eq:2}
  \tilde{S}_\pm=\bar{S}_\pm-\frac{\hbar^2}{m_0}\frac{2\sqrt{3}}{3}k_{\pm}[\kappa,\hat{k}_z],~~~~~~
  \hat{C} = \frac{\hbar^2}{m_0}k_{-}[\kappa,\hat{k}_z],~~~~~~k_\pm = k_x \pm ik_y,~~~~~~\hat{k}_z =-i\frac{\partial}{\partial z}.
\end{equation}
Here, $[\hat{A},\hat{B}]=\hat{A}\hat{B}-\hat{B}\hat{A}$ is the commutator, $\{\hat{A},\hat{B}\}=\hat{A}\hat{B}+\hat{B}\hat{A}$ is the anticommutator for the operators $\hat{A}$ and $\hat{B}$; $\tilde{P}$ is the Kane momentum matrix element, $\tilde{P}^2=\hbar^2 E_P/2m_0$; $E_{c}(z)$ and $E_{v}(z)$ are the conduction and valence band edges, respectively; $\Delta(z)$ is the spin orbit energy; $\gamma_1$, $\gamma_2$, $\gamma_3$, $\kappa$ and $F$ describe the interaction with the remote bands, not considered in the Hamiltonian. It is assumed that the $z$ axis coincides with the crystallographic direction (0$lk$), while the $x$ and $y$ axes correspond to directions (100) and (0$k\bar{l}$), respectively. The terms $\hat{T}^{(\epsilon)}$, $\hat{U}^{(\epsilon)}$, $\hat{V}^{(\epsilon)}$, $\hat{R}^{(\epsilon)}$, $\hat{S}_{\pm}^{(\epsilon)}$ in Eqs~\eqref{eq:2}, resulting from lattice-mismatch strain, are written as follows:
\begin{equation*}
\hat{T}^{(\epsilon)} = a_c(2\epsilon_{xx}+\epsilon_{zz}),~~~~~~~~~~~
\hat{U}^{(\epsilon)} = a_v(2\epsilon_{xx}+\epsilon_{zz}),
\end{equation*}
\begin{equation*}
\hat{V}^{(\epsilon)}=b(\epsilon_{xx}-\epsilon_{zz})-\frac{1}{4}\sin^2(2\varphi)\left(3b-\sqrt{3}d\right)(\epsilon_{xx}-\epsilon_{zz})+\frac{1}{4}\sin(4\varphi)\left(3b-\sqrt{3}d\right)\epsilon_{yz},
\end{equation*}
\begin{equation*}
\hat{R}^{(\epsilon)}=-\frac{1}{4}\sin^2(2\varphi)\left(\sqrt{3}b-d\right)(\epsilon_{xx}-\epsilon_{zz})+\frac{1}{4}\sin(4\varphi)\left(\sqrt{3}b-d\right)\epsilon_{yz},
\end{equation*}
\begin{equation}
\label{eq:3}
\hat{S}_{\pm}^{(\epsilon)}=\mp \frac{i}{4}\sin(4\varphi)\left(d-\sqrt{3}b\right)(\epsilon_{xx}-\epsilon_{zz})\pm i\left(d-\frac{\sqrt{3}}{2}b\right)\epsilon_{yz}\pm i\sin^2(2\varphi)\left(d-\sqrt{3}b\right)\epsilon_{yz},
\end{equation}
where $a_c$ and $a_v$ are the hydrostatic deformation potentials, while $b$ and $d$ are the uniaxial deformation potentials;  $\epsilon_{xx}=\epsilon_{yy}$, $\epsilon_{zz}$ and $\epsilon_{yz}$ are non-zero components of the strain tensor. From the condition of zero external stress along the (0$lk$) direction we get the relation between $\epsilon_{xx}$, $\epsilon_{zz}$ and $\epsilon_{yz}$:\cite{w27a}
\begin{equation*}
\epsilon_{xx}=\frac{a_0-a_L}{a_L},
\end{equation*}
\begin{equation*}
\epsilon_{zz}=\frac{c_{11}^2+2 c_{11}\left(c_{12}-c_{44}\right)+c_{12}\left(-3 c_{12}+10 c_{44}\right)-\left(c_{11}+3 c_{12}\right)\left(c_{11}-c_{12}-2 c_{44}\right)\cos(4\varphi)}{-c_{11}^2-6c_{11}c_{44}+c_{12}\left(c_{12}+2c_{44}\right)+\left(c_{11}+ c_{12}\right)\left(c_{11}-c_{12}-2 c_{44}\right)\cos(4\varphi)}\epsilon_{xx},
\end{equation*}
\begin{equation}
\label{eq:4}
\epsilon_{yz}=-\frac{\left(c_{11}+2 c_{12}\right)\left(c_{11}-c_{12}-2 c_{44}\right)\sin(4\varphi)}{-c_{11}^2-6c_{11}c_{44}+c_{12}\left(c_{12}+2c_{44}\right)+\left(c_{11}+ c_{12}\right)\left(c_{11}-c_{12}-2 c_{44}\right)\cos(4\varphi)}\epsilon_{xx},
\end{equation}
where $c_{ij}$ are the elastic constants in each layer, $a_L$ and $a_0$ are the lattice parameters of the given layer and the buffer, respectively.

Assuming translation invariance in the $xy$ plane, the envelope function $F_i(\textbf{r})$ for $u_i(\textbf{r})$ Bloch amplitude\cite{w21} can be represented as
\begin{equation}
\label{eq:10}
F_i(\textbf{r})=\exp\left(ik_{x}x+ik_{y}y\right)f_{i}(z),
\end{equation}
where $k_x$ and $k_y$ are the wave vector components in the QW plane. As a result, Schr\"{o}dinger equation with the 8-band \textbf{k$\cdot$p} Hamiltonian is reduced to the following system of differential equations:
\begin{equation}
\label{eq:11}
\sum_{j=1}^{8}(\hat{H}_{\textbf{k}}^{8\times 8})_{ij}f_i(z)=E_{n_z}(k_x, k_y)f_i(z),
\end{equation}
where $n_z$ is the electronic subband index. To solve this system, the functions $f_i(z)$ are expanded in terms of the complete basis set $\{\eta_n\}$ of plane waves:
\begin{equation}
\label{eq:12}
f_i(z)=\dfrac{1}{\sqrt{L_z}}\sum_{n=-N}^{N}C_{i}^{(n)}\exp(ik_nz),
\end{equation}
where $k_n=2\pi n/L_z$ and $L_z$ is the total width of QW structure in $z$ direction (in this work, $L_z=2L_{CdHgTe}+d$, where $L_{CdHgTe}=$ 30~nm). In our calculations, $N$ defines the accuracy of the solution of the eigenvalue problem, $N=$ 90 is good to get convergent results with precision higher than 0.5~\%.

The expansion in Eq.~\eqref{eq:12} leads to a matrix representation of the eigenvalue problem, where the eigenvectors with components $C_{i}^{(n)}$ and the corresponding eigenvalues are obtained by diagonalization of matrix $\langle\eta_n|(\hat{H}_{\textbf{k}}^{8\times 8})_{ij}|\eta_{n'}\rangle$. By using the plane-wave basis, the matrix elements $\langle\eta_n|K(z)|\eta_{n'}\rangle$, $\langle\eta_n|\partial_{z}K(z)|\eta_{n'}\rangle$, and $\langle\eta_n|\partial_{z}K(z)\partial_{z}|\eta_{n'}\rangle$ can be calculated analytically, where $K(z)$ is an arbitrary polynomial for each of the QW layers.

With the basis expansion method, through the eigenvectors $\textbf{C}$ in Eq.~\eqref{eq:12}, we can easily classify the levels. For electronic subband $n_z$, we define the relative contribution to this level from the basis states in the set $I$:
\begin{equation}
\label{eq:13}
d_I(k_x,k_y)=\sum_{n=-N}^{N}\sum_{i\in I}\left|C_{i}^{(n)}(E_{n_z},k_x, k_y)\right|^2,
\end{equation}
where $d_I(k_x,k_y)$ is normalized such that if we include all the states in the set $I$, then $d_I(k_x,k_y)=1$.

In this work, we calculate $d_{e}$ for the contribution from the $|\Gamma_6, \pm1/2\rangle$ states, $d_{lh}$ for the contribution from the $|\Gamma_8, \pm1/2\rangle$ states, $d_{so}$ for the contribution from the $|\Gamma_7, \pm1/2\rangle$ states and $d_{hh}$ for the contribution from the $|\Gamma_8, \pm3/2\rangle$ states. For example, to calculate $d_{hh}$ from Eq.~\eqref{eq:13}, we let $I$ contain $i=$ 3, 6. It is clear that $d_{e}+d_{lh}+d_{so}+d_{hh}=1$ at any values of $\textbf{k}$. We classify electronic subbands in HgTe QW as electron-like or hole-like levels by comparing the value of $d_{e}+d_{lh}+d_{so}$ with $d_{hh}$. The given subband is the hole-like level if $d_{hh}>d_{e}+d_{lh}+d_{so}$ at $\textbf{k}=0$. Otherwise, the subbands are classified as electron-like, light-hole-like or spin-off-like levels, according to the dominant component in the sum $d_{e}+d_{lh}+d_{so}$ at $\textbf{k}=0$.

\subsection{Calculation of Landau levels}

To calculate the energy levels in perpendicular magnetic field $\textbf{B}$=(0,~0,~$B$) we use a Peierls substitution
\begin{equation*}
k_{x}=-i\frac{\partial}{\partial x}+\frac{e}{\hbar c}A_{x},
\end{equation*}
\begin{equation}
\label{eq:14}
k_{y}=-i\frac{\partial}{\partial y}+\frac{e}{\hbar c}A_{y}
\end{equation}
and introduce the ladder operators $b^{+}$ and $b$:
\begin{equation*}
b^{+}=\frac{a_{B}}{\sqrt{2}}k_{+},
\end{equation*}
\begin{equation*}
\label{eq:15}
b=\frac{a_{B}}{\sqrt{2}}k_{-},
\end{equation*}
where $a_{B}$ is the magnetic length ($a_{B}^{2} =c\hbar/eB$), $e>0$ is the elementary charge and $\textbf{A}$ is the magnetic vector potential in Landau gauge $\textbf{A}$= (0, $Bx$,0).

In addition to $\hat{H}_{\textbf{k}}^{8\times 8}$, one needs also to take into account the Zeeman term $H_z$, which has the form
\begin{equation}
H_z =\mu_{B}B\begin{pmatrix}
1 & 0 & 0 & 0 & 0 & 0 & 0 & 0 \\
0 & -1 & 0 & 0 & 0 & 0 & 0 & 0 \\
0 & 0 &-3\kappa & 0 & 0 & 0 & 0 & 0 \\
0 & 0 & 0 & -\kappa & 0 & 0 & -\sqrt{2}\kappa & 0 \\
0 & 0 & 0 & 0 & \kappa & 0 & 0 & -\sqrt{2}\kappa \\
0 & 0 & 0 & 0 & 0 & 3\kappa & 0 & 0 \\
0 & 0 & 0 & -\sqrt{2}\kappa & 0  & 0 & -2\kappa & 0  \\
0 & 0 & 0 & 0 & -\sqrt{2}\kappa & 0 & 0 & 2\kappa
\end{pmatrix}.
\end{equation}

To calculate LLs, we use so-called the axial approximation. Within this approximation we keep the in-plane rotation symmetry by omitting the warping terms in $T$, $U$, $V$, $R$, $\bar{S}_\pm$ and   $\tilde{S}_\pm$, which now are written as follows:
\begin{equation*}
  T = E_c + \frac{\hbar^2}{2m_0}\left\{
    (2F+1)\dfrac{{k}_{+}{k}_{-}+{k}_{-}{k}_{+}}{2}
    + \hat{k}_z (2F+1) \hat{k}_z\right\}+T^{(\epsilon)},
\end{equation*}
\begin{equation*}
  U = E_v - \frac{\hbar^2}{2m_0}\left\{
    \gamma_1\dfrac{{k}_{+}{k}_{-}+{k}_{-}{k}_{+}}{2}
    + \hat{k}_z \gamma_1 \hat{k}_z\right\}+U^{(\epsilon)};
\end{equation*}
\begin{multline*}
  V = -\frac{\hbar^2}{2m_0}\Biggl\{
    \gamma_2 \dfrac{{k}_{+}{k}_{-}+{k}_{-}{k}_{+}}{2}-2\hat{k}_{z}\gamma_{2}\hat{k}_{z}-\frac{3}{4}\sin^2(2\varphi)\left(\gamma_2-\gamma_3\right) \dfrac{{k}_{+}{k}_{-}+{k}_{-}{k}_{+}}{2}+ \\
    + \frac{3}{2}\sin^2(2\varphi)\hat{k}_z \left(\gamma_2-\gamma_3\right) \hat{k}_z  \Biggr\}+V^{(\epsilon)};
\end{multline*}
\begin{equation*}
  R = \frac{\hbar^2}{2m_0}\sqrt{3}{k}_{-}^2\Biggl\{    \dfrac{\gamma_2+\gamma_3}{2}-\frac{3}{8}\sin^2(2\varphi)\left(\gamma_2-\gamma_3\right)\Biggr\};
\end{equation*}
\begin{equation*}
  \bar{S}_\pm = -\frac{\hbar^2}{2m_0}\sqrt{3}k_{\pm}\Biggl\{
   \{\gamma_3,\hat{k}_z\}+[\kappa,\hat{k}_z]+ \frac{1}{2}\sin^2(2\varphi) \Bigl\{\gamma_2-\gamma_3,\hat{k}_z\Bigr\}
  \Biggr\};
\end{equation*}
\begin{equation}
\label{eq:16}
  \tilde{S}_\pm =-\frac{\hbar^2}{2m_0}\sqrt{3}k_{\pm}\Biggl\{
    \{\gamma_3,\hat{k}_z\}- \frac{1}{3}[\kappa,\hat{k}_z]+\frac{1}{2}\sin^2(2\varphi) \Bigl\{\gamma_2-\gamma_3,\hat{k}_z\Bigr\}\Biggr\}.
\end{equation}
We note that expressions, written above, contain the right order of the operators $k_{+}$ and $k_{-}$ in the presence of magnetic field.

In the axial approximation, the total wave function can be written as

\begin{equation*}
\Psi_{n_{z},n,\tilde{k}}^{(i)}=
\begin{pmatrix}
c_{1}^{(i)}(z,n_{z},n)|n,\tilde{k}\rangle \\[3pt]
c_{2}^{(i)}(z,n_{z},n)|n+1,\tilde{k}\rangle \\[3pt]
c_{3}^{(i)}(z,n_{z},n)|n-1,\tilde{k}\rangle \\[3pt]
c_{4}^{(i)}(z,n_{z},n)|n,\tilde{k}\rangle \\[3pt]
c_{5}^{(i)}(z,n_{z},n)|n+1,\tilde{k}\rangle \\[2pt]
c_{6}^{(i)}(z,n_{z},n)|n+2,\tilde{k}\rangle \\[3pt]
c_{7}^{(i)}(z,n_{z},n)|n,\tilde{k}\rangle \\[3pt]
c_{8}^{(i)}(z,n_{z},n)|n+1,\tilde{k}\rangle
\end{pmatrix},
\end{equation*}
\begin{equation*}
|n,\tilde{k}\rangle=
\begin{cases}
0,~n<0,\\[3pt]
\dfrac{\exp\left(i\tilde{k}y\right)}{\sqrt{2^{n}n!\sqrt{\pi}a_{B}L_{y}}}H_{n}\left(\dfrac{\tilde{x}}{a_{B}}\right)\exp\left(-\dfrac{\tilde{x}^{2}}{2a_{B}^{2}}\right),~n\geq0.
\end{cases}
\end{equation*}
\begin{equation}
\label{eq:17}
\tilde{x}=x-\tilde{k}a_{B}^2,
\end{equation}
where $L_{y}$ is the sample size along the $y$ axis, $H_{n}$ are the Hermitian polynomials with number $n$ ($n$ is also the Landau level index and the eigenvalue of the operator $b^{+}b$), $\tilde{k}$ is the wave vector projection onto the $y$ axis.

For $n=-2$, there is one-component wave function, which is not mixed with other LLs at $n>-2$ and is formed by heavy-hole states $|\Gamma_8,-3/2\rangle$ only:
\begin{equation*}
\Psi_{n_{z},-2,\tilde{k}}^{(i)}=
\begin{pmatrix}
0 \\[3pt]
0 \\[3pt]
0 \\[3pt]
0 \\[3pt]
0 \\[2pt]
c_{6}^{(i)}(z,n_{z},-2)|0,\tilde{k}\rangle \\[3pt]
0 \\[3pt]
0
\end{pmatrix}.
\end{equation*}
We note that this LL, together with one of the characteristic solutions for $n=0$, represents so-called zero-mode LLs, which are identified within a simplified approach of the Dirac-type Hamiltonian,\cite{w5,w12} mentioned in the main text of this paper. To solve the Schr\"{o}dinger equation under magnetic field, we also expand functions $c_{i}(z,n,n_{z})$, $i=$1,…8, by a series of plane waves, as it is done in the absence of magnetic field.

\begin{figure}
\includegraphics [width=0.95\columnwidth, keepaspectratio] {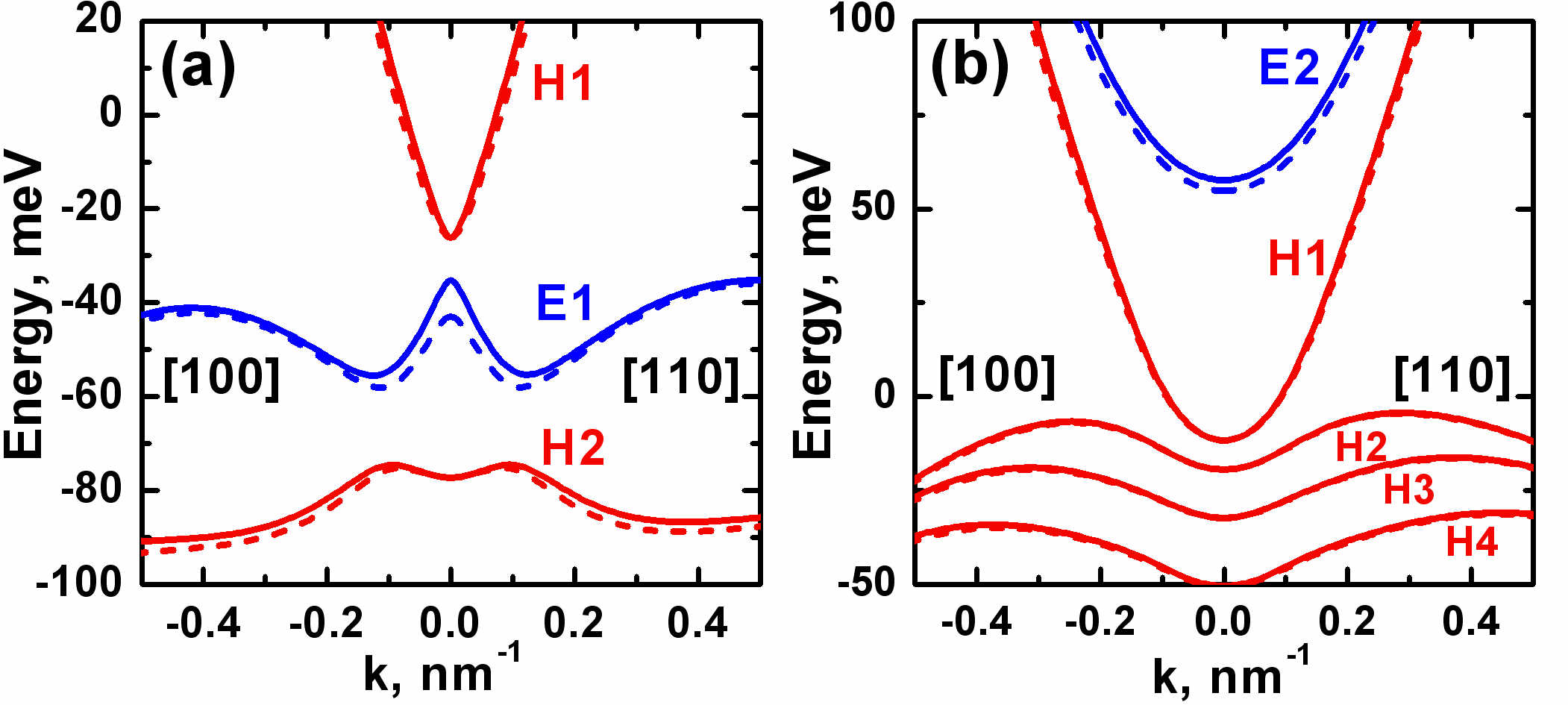} 
\caption{\label{Fig:1SM} (Color online) Band structure of HgTe/Cd$_{0.7}$Hg$_{0.3}$Te QW, grown on (001) CdTe buffer at $T=$ 0 K and $P=$ 0 kBar for different QW width: (a) 8 nm (TI phase) and (b) 20 nm (SM phase). The electron-like subbands are shown in blue, the red curves correspond to the heavy-hole-like subbands. In the panel (b), the E1 subband lies significantly lower in energy. Solid curves are the calculations within the 8-band \textbf{k$\cdot$p} Hamiltonian for the $\Gamma_6$, $\Gamma_8$ and $\Gamma_7$ bands, while the dashed curves conform to the 6-band \textbf{k$\cdot$p} Hamiltonian, in which the coupling with the $\Gamma_7$ band is ignored.}
\end{figure}

\subsection{Effect of interaction with the $\Gamma_7$ band on the band structure of HgTe/Cd(Hg)Te QWs}

Often, the $\Gamma_7$ band is ignored into band structure calculations, assuming that it has negligible effects. This is possible in the limit of large $\Delta$. In this case, the 8-band \textbf{k$\cdot$p} Hamiltonian can be easily projected on the subspace, orthogonal to the $\Gamma_7$ band. The projection is done by simply eliminating the seventh and the eighth row and column of the matrix in Eq.~\ref{eq:1}. Such reduced Hamiltonian, written only for the $\Gamma_6$ and $\Gamma_8$ bands, is called as the 6-band \textbf{k$\cdot$p} Hamiltonian.\cite{w27a}

Fig.~\ref{Fig:1SM} shows the band structure of 8 and 20 nm thick (001)-oriented HgTe QWs, calculated by using the 6-band and the 8-band \textbf{k$\cdot$p} Hamiltonians. One can see that the $\Gamma_7$ band effect is relevant for the electron-like states but it does not affect positions of heavy-hole-like subbands at $k_{x,y} = 0$. One can also see from the 8-band \textbf{k$\cdot$p} Hamiltonian for (001)-oriented QWs ($\varphi=0$) that the heavy-hole-like subbands, formed by $|\Gamma_8,\pm3/2\rangle$ states, at $k_{x,y} = 0$ are decoupled from the electron-like states, formed by superposition of $|\Gamma_6,\pm1/2\rangle$, $|\Gamma_8,\pm1/2\rangle$ and $|\Gamma_7,\pm1/2\rangle$ states. Therefore, the $\Gamma_7$ band does not affect the heavy-hole-like levels in the QW at zero quasimomentum in the plane. If $k_{x,y} \neq 0$, the electron-like and hole-like states are mixed, and the effect arises even for the H1, H2, H3, H4 subbands.

Actually, one can not accurately use the 6-band \textbf{k$\cdot$p} Hamiltonian for thin HgTe QWs because the position of the $\Gamma_8$ band in CdTe layer exceeds that of the $\Gamma_7$ band in HgTe of only 510 meV at zero pressure at temperature. The latter is comparable with the Valence Band Offset, arising in HgTe/CdTe interface. Therefore, in thin QWs, for which effect of mixing between the $\Gamma_8$ and $\Gamma_7$ bands from different layers is strong, the 6-band \textbf{k$\cdot$p} Hamiltonian leads to significant deviation from the results obtained within the 8-band model (see Fig.~\ref{Fig:1}a). However, for the thick QWs, for which the role of the interface mixing decreases with QW width, the deviation vanishes.


%
\end{document}